\documentclass{Interspeech2024}

\interspeechcameraready

\title{Speaker-Smoothed kNN Speaker Adaptation for End-to-End ASR}

\name[]{Shaojun}{Li}
\name[]{Daimeng}{Wei}
\name[]{Hengchao}{Shang}
\name[]{Jiaxin}{GUO}
\name[]{ZongYao}{LI}
\name[]{Zhanglin}{Wu}
\name[]{Zhiqiang}{Rao}
\name[]{Yuanchang}{Luo}
\name[]{Xianghui}{He}
\name[]{Hao}{Yang}

\address{
  Huawei Translation Services Center}
\email{\{lishaojun18,weidaimeng,shanghengchao,guojiaxin1,lizongyao,wuzhanglin2,raozhiqiang,
luoyuanchang1,hexianghui,yanghao30\}@huawei.com}

\keywords{speech recognition, speaker adaptation, k-nearest neighbors}

\begin{document}

\maketitle

\begin{abstract}
    
Despite recent improvements in End-to-End Automatic Speech Recognition (E2E ASR) systems, the performance can degrade due to vocal characteristic mismatches between training and testing data, particularly with limited target speaker adaptation data. We propose a novel speaker adaptation approach Speaker-Smoothed kNN that leverages k-Nearest Neighbors (kNN) retrieval techniques to improve model output by finding correctly pronounced tokens from its pre-built datastore during the decoding phase. Moreover, we utilize x-vector to dynamically adjust kNN interpolation parameters for data sparsity issue. This approach was validated using KeSpeech and MagicData corpora under in-domain and all-domain settings. Our method consistently performs comparably to fine-tuning without the associated performance degradation during speaker changes. Furthermore, in the all-domain setting, our method achieves state-of-the-art results, reducing the CER in both single speaker and multi-speaker test scenarios.
\end{abstract}

\section{Introduction}

Recently, end-to-end automatic speech recognition (E2E ASR) systems have demonstrated considerable performance improvements, aided by extensive training data amassed from a variety of speakers \cite{Watanabe_Hori_Kim_Hershey_Hayashi_2017,sainath2020streaming}. Despite these advancements, the performance of E2E systems can plummet dramatically when confronted with significant voice character mismatch between training and testing conditions. In response to this issue, speaker adaptation algorithms seek to rectify the aforementioned mismatch by tailoring the ASR model to fit the specific characteristics of the target speaker.

The biggest challenge of speaker adaptation is that the adaptation data amount from the target speaker is usually very small. There are two types of methods to address such a challenge. 
The first type is model-based method, it studies how to effectively utilize few-shot speaker data. Some methods directly fine-tune the parameters of the pre-trained model \cite{Huang_Ye_Li_Gong_2021}, or train the model from speaker-independent parameters \cite{7078569,Xie2020BayesianLF}. To alleviate the overfitting problem caused by the limited adaptation data, the L2 norm \cite{Liao_2013}, Kullback–Leibler divergence \cite{Yu_Yao_Su_Li_Seide_2013,Meng2019SpeakerAF} are introduced to regularize the adapted model. There are also many approaches utilizing data augmentation, for example, generating speaker data via TTS to augment the model \cite{yang2023text}. However, most of these methods potentially suffer from significant performance drops when the amount of adaptation data decreases.

The second type is feature-based method, in which auxiliary speaker embeddings, such as i-vector \cite{5545402} and x-vector \cite{Snyder_Garcia-Romero_Sell_Povey_Khudanpur_2018}, are fed into an ASR model along with speech features.
In this way, the model is working on the acoustic features, i.e. either by normalizing acoustic features to be speaker-independent \cite{feature2,feature3} or by introducing additional speaker related knowledge (e.g., i-vector) to adapt the acoustic model \cite{feature5,feature7}. A summary vector of each utterance can be trained to replace speaker i-vector \cite{feature10}. To adapt to acoustic variability \cite{feature11}, shifting and scaling parameters are added in the layer-normalization layer. Recently, the attention mechanism is introduced to speaker-aware training (SAT). Typically, a speaker-aware persistent memory is incorporated to the transformer based ASR model \cite{sarı2020unsupervised,zhao20b_interspeech}. These memories performance always hit a bottleneck with the limited capacity of tens or hundreds of speaker embedding search space.

Considering the large size of E2E ASR training set, there may be utterances of similar voice characters as that of the target speaker. Thus, it is reasonable to utilize the similar utterances to be a supplement for target speaker data in the adaptation process. Motivated by this idea, we introduce a kNN retrieval techniques to ASR model which initially proposed for regulating autoregressive decoding (AD) of language model \cite{Khandelwal2019GeneralizationTM} and machine translation \cite{Khandelwal2020NearestNM}. We expect this kNN classifier to find the correctly pronounced token from its pre-built datastore at each decoding to correct the model's erroneous output. \cite{10447075} modified kNN to apply to frame-level CTC decoding, but lacked further research on AD and the risk of false recall when data is sparse in kNN \cite{jiang2021learning}. Therefore, we firstly explore the effectiveness of kNN token-level representation. It was found that there may be some problems when speaker adaptation is used directly on kNN, because kNN has the fixed interpolation parameters, $T$ and $\lambda$. To solve this problem, we propose a Speaker-Smoothed kNN network, which uses extra information such as x-vector to dynamically adjust the interpolation ratio. \cite{gu23_interspeech} uses few-shot speaker data to extract similar pronunciations from training data, which is similar to our idea. However, this training method only uses utterance level information and re-training is required when the speaker is switched.

We test proposed Speaker-Smoothed kNN using KeSpeech and MagicData copora with in-domain and all-domain settings. In the in-domain setting, our method performs close to fine-tuning without experiencing performance degradation like fine-tune and kNN method when the speaker changes. In the all-domain setting, our method achieves sota in both single speaker and speaker change scenarios, reduce 12.35 and 24.68 CER on the single speaker and multi-speaker test sets respectively.

\section{Background}

Here we give a brief introduction of kNN method. kNN uses token-level context retrieval to enhance the quality of a pre-trained neural machine translation (NMT) model \cite{vaswani2023attention}. This method includes two steps:

\noindent \textbf{Datastore Construction} The datastore is a translation memory which converts bilingual sentence pairs into a set of key-value pairs. Given a reference corpus $(x,y) \in (\mathcal{X},\mathcal{Y})$, the pre-trained NMT model generates the context representation $f_{\theta}(x,y_{<t})$ at each timestep $t$.
Then we collect the output hidden state $f_{\theta}(x,y_{<t})$ as key and $y_t$ as value to construct the whole datastore $(\mathcal{K},\mathcal{V})$:

\begin{equation}
    (\mathcal{K}, \mathcal{V}) = \bigcup_{(x,y)\in (\mathcal{X}, \mathcal{Y})}\{(f_{\theta}(x,y_{<t}), y_t), \forall y_t \in y\}
\label{equ:ds-build}
\end{equation}

\noindent \textbf{Inference with kNN Retrieval} At the $t$-th decoding step, given the already generated words $\hat y_{<t}$, the current context representation $f_{\theta}(x,\hat y_{<t})$ is leveraged to generate a retrieval distribution $p_{k\mathrm{NN}}(y_t|x,\hat y_{<t})$ over the entire vocabulary:
\begin{equation}
    p_{k\mathrm{NN}}(y_t|x,\hat y_{<t}) \propto  
     \sum_{(h_i,v_i) \in N_t}  \mathbb{I}_{y_t=v_i} \mathrm{exp} (\frac{-d(h_i, f_{\theta}(x,\hat y_{<t}))^2}{T})
\label{equ:knn-dist}
\end{equation}

where the $d(.,.)$ stands for Euclidean distance function and $T$ is the temperature to control the sharpness of softmax function.
The final prediction distribution enhances vanilla NMT distribution $p_{\textrm{NMT}}$ with the retrieval distribution $p_{k\mathrm{NN}}$, and it is formally calculated as:
\begin{equation}
        p(y_t|x,\hat y_{<t}) = \lambda p_{k\mathrm{NN}}(y_t|x,\hat y_{<t}) 
         + (1-\lambda)  p_{\mathrm{NMT}}(y_t|x,\hat y_{<t})
\label{equ:p_knn}
\end{equation}
$T$ and $\lambda$ above are tuned interpolation coefficients.

\section{Method}
\begin{figure}[t]
  \centering
  \includegraphics[width=\linewidth]{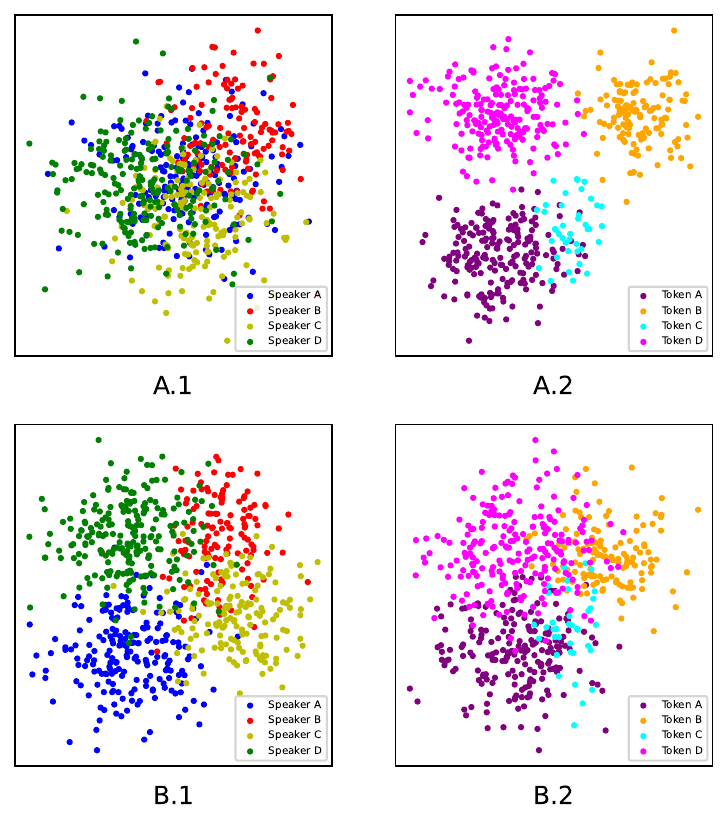}
  \caption{t-SNE result of the kNN datastore representation. A.x samples representation from the same subdialect, and B.x samples from different subdialects. Dots of different colors indicate different speaker or token clusters. }
  \label{fig:scatter}
\vspace{-0.3cm} %
\end{figure}

\subsection{Preliminary Study}
Given the limited research on the application of kNN to ASR and speaker adaptation, we initiated a preliminary study to demonstrate the potential of kNN in ASR. Utilizing Whisper, we obtained token-level representations denoted as $h_i$. We then sampled different speaker kNN representations across three dimensions, they are subdialect, speaker and token. The t-SNE results on $h_i$ embeddings from Whisper are illustrated in Figure \ref{fig:scatter}. A.x samples derive from the same subdialect, where A.1 samples the same tokens with varying colors representing different speakers, while A.2 includes multiple speakers, with different color schemes indicating different tokens. B.x settings, on the other hand, sample from different subdialects, choosing up to 200 representations for four tokens or speakers.

The A.x samples drawn from the same subdialect exhibit considerable overlap among speakers within the same token cluster (A.1), and a clear demarcation exists between different token clusters (A.2). These results align with our expectations and benefit our kNN retrieval. It is important to note the evident sparseness of the speaker adaptation data, as reflected in token cluster A.2 (token C), as the use of fixed temperatures and weights by kNN for interpolation increases the chance of error tokens.

B.x samples, collected from different subdialects, display overlaps between clusters in B.1 and boundaries between clusters in B.2, as anticipated. However, an unfortunate overlap is found in token C of B.2. We hypothesize that this overlap may be attributed to similar pronunciations of different tokens by speakers from diverse subdialects, in addition to data sparseness.

The above-mentioned overlaps, which kNN classification cannot distinguish effectively, are largely related to speaker accents. Therefore, the incorporation of speaker embedding into kNN classification may prove beneficial in distinguishing between these overlaps, thereby enhancing the accuracy of our approach.

\subsection{Speaker-Smoothed kNN Network}

\begin{figure*}[t]
    \small
    \centering
    \includegraphics[width = 17cm]{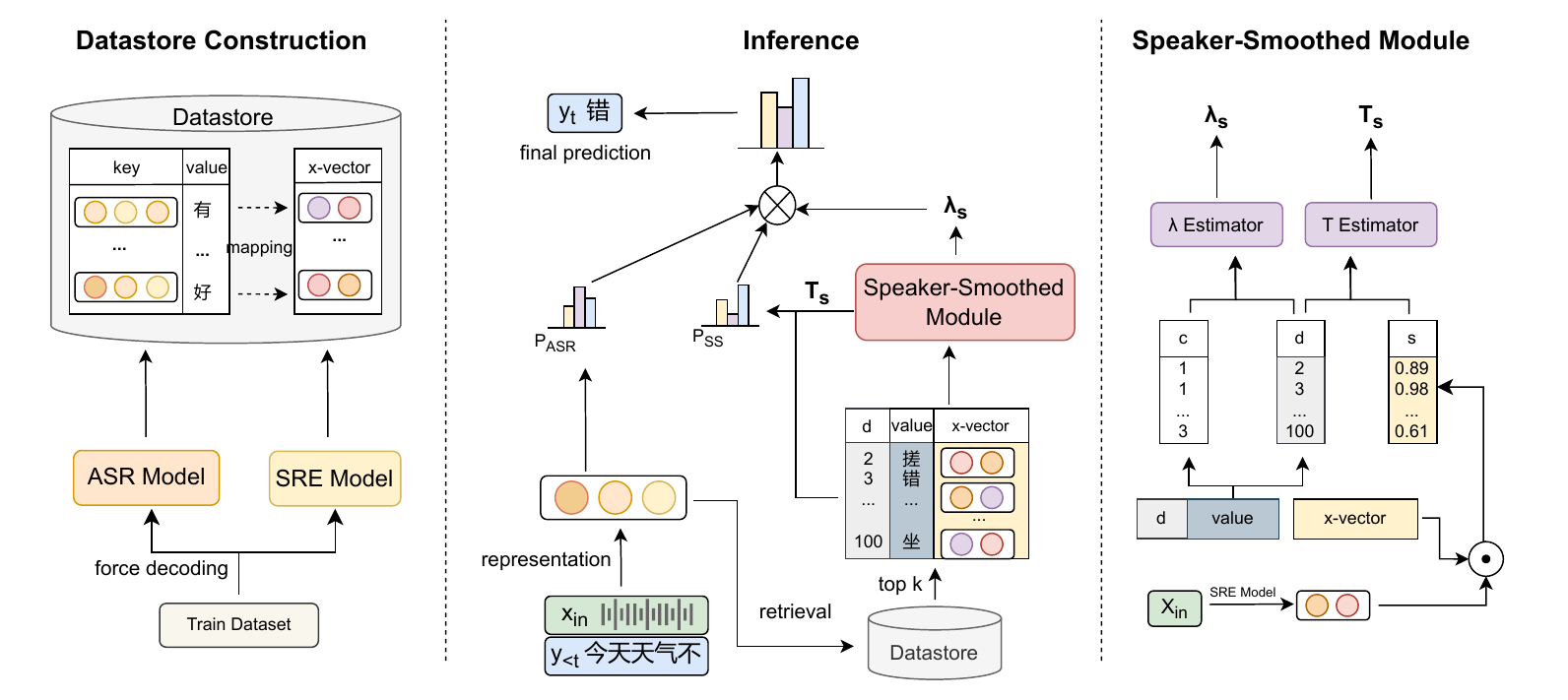}
    \caption{An overview of our proposed Speaker-Smoothed kNN framework}
    \label{fig:arch}
\end{figure*}

Speaker recognition (SRE) models have been demonstrated to effectively distinguish between speakers \cite{8461375,bai2021speaker}. Speaker embeddings, derived from SRE models, have also been deployed directly for speaker adaptation \cite{sarı2020unsupervised,zhao20b_interspeech}. To address the challenges posed by sparsity and error overlap identified in our preliminary study, we now incorporate x-vector, a type of speaker information, into the kNN approach.

Figure \ref{fig:arch} displays our process, starting with datastore construction as per Section 2. We use the SRE model to create an utterance-level x-vector, which we pair with its kNN value in a mapping relationship. Kster's work \cite{jiang2021learning} inspires us to design two estimators for variables $T$ and $\lambda$ dynamically. Unlike Kster, we simplify our network settings according to \cite{Dai2023SimpleAS} and integrate x-vector. Noting the sparsity of our speaker data, we heed findings from \cite{Zheng2021AdaptiveNN}, asserting networks utilizing distance and distinct token information yield better generalization results.

Concretely, at the $t$-th decoding step, we first retrieve $K$ neighbors from the datastore, and consider the powers of 2 as the choices of $k$ for simplicity. For each neighbor $(h_i, v_i, e_i)$, where $e_i$ denoted the mapping x-vectors, we calculate the distance from the current context representation $d_i=d(h_i, f(x, \hat{y}_{<t}))$, as well as the count of distinct values in top $i$ neighbors $c_i$.
Denote $d=(d_1,...,d_K)$ as distances and $c=(c_1,...,c_K)$ as counts of values for all retrieved neighbors. Finally, we use input $x$ to generate x-vector and calculate the dot product with $e$, here, denote $s=(s_1,...,s_K)$ as the x-vector similarity.

For dynamic distribution temperature $T$ denoted as $T_s$, we concatenate $d$ and $s$ as input features to the $T$ estimator network. From the preliminary study, the above operations are performed in the hope that the x-vector will help distinguish overlapping issues.

\begin{equation}
    T_s = \exp(\mathbf{W}_{1}[d;s] + \mathbf{b}_{1})
\end{equation}

For dynamic $\lambda$ denoted as $\lambda_s$, the mixing weight $\lambda_s$ is computed by a $d$ and $c$ as inputs where $[\mathbf{W}_2;\mathbf{b}_2; \mathbf{W}_3; \mathbf{b}_3]$ are trainable parameters. These distance information can deal with the sparse problem from the preliminary study.

\begin{equation}
    \mathbf{\lambda_s} = \text{sigmoid}(\mathbf{W}_{3}\text{ReLU}(\mathbf{W}_{2}[d;c] + \mathbf{b}_{2}) + \mathbf{b}_3) \\
\end{equation}

It's noteworthy that our technique, in theory, can be employed with any autoregressive E2E ASR model, including RNN-T or Conformer. Once our network has been trained, it enables hot update in the speaker datastore during testing, thereby supporting on-the-fly adaptation.

\subsection{Prediction and Training}
In prediction phrase, we use our Speaker-Smoothed network to generate dynamic $T_s$ and $\lambda_s$ for each untranscribed token to utilize different T and $\lambda$.  That is to say, we replace the fixed $T$ in Equation \ref{equ:knn-dist} and $\lambda$ in Equation \ref{equ:p_knn} by $\lambda_s$ and $T_s$ varies at each predicting $\hat y_t$. Here $p_{ASR}$ denotes ASR distribution and $p_{SS}$ denotes Speaker-Smoothed kNN distribution 

\begin{equation}
        p(y_t|x,\hat y_{<t}) = \lambda_s p_{ASR}(y_t|x,\hat y_{<t}) 
         + (1-\lambda_s)  p_{\mathrm{SS}}(y_t|x,\hat y_{<t})
\label{equ:p_smooth}
\end{equation}

For training, we fix the pre-trained ASR model and only optimize the Speaker-Smoothed kNN network by minimizing the cross entropy loss following Equation~(\ref{equ:p_smooth}), which could be very efficient by only utilizing hundreds of training samples.

\section{Experiment}
\begin{table}[tbp]
\centering
\small
\tabcolsep 8pt
\begin{tabular}{l|cc}
\toprule

Subdialect & \#Hours   & \#Million Tokens  \\
\midrule
All-Domain              & 860    &  11               \\    
\midrule
\multicolumn{3}{l}{In-Domain} \\
\midrule
\textit{Zhongyuan}      & 84  & 1.6            \\
\textit{Southwestern}   & 75  & 1.4         \\
\textit{Ji-Lu}          & 59  &1.1     \\
\textit{Jiang-Huai}     & 46  &0.9      \\
\bottomrule
\end{tabular}
\caption{Number hours and tokens of in-domain and all-domain data collection in KeSpeech.}
\label{tab:data_static}
\vspace{-0.5cm} %
\end{table}

\subsection{Setup}

We leverage the Whisper-medium pre-trained model for tracking speaker mismatches. This model demonstrates impressive efficacy on Mandarin Chinese tests. A pre-trained speaker recognition model aids \cite{Landini_Profant_Diez_Burget_2020} in extracting embeddings for similarity assessments and calculating variable $s$ as per Section 3.2.

For training and kNN datastore construction, the KeSpeech \footnote{\url{https://github.com/KeSpeech/KeSpeech}} corpora is employed. We perform two categories of experiments based on datasets: (1) In-domain setting, we train exclusively with a dataset composed of a single subdialect; and (2) All-domain setting, utilizing the entire KeSpeech corpus - an open-source dataset of Mandarin and eight subdialects. We deploy phase-1 of KeSpeech, offering 895 hours of data from 34 cities across China. Table \ref{tab:data_static} details training data and datastore size.

Our evaluation employs two test sets based on speaker variation: (1) Single speaker test set. For this, we utilize two open-source datasets from the MagicData \footnote{\url{https://magichub.com/datasets}}: the Sichuan and Zhengzhou corpora, each representing Southwestern and Zhongyuan Mandarin dialects, respectively. These single speaker adaptation test sets consist of 20 speakers, each providing approximately 30 minutes of speech. The test sets are allocated following \cite{gu23_interspeech}, allocating 10 minutes for adaptation and 20 minutes for testing. (2) Multi-speaker test set. Here, we leverage the entire KeSpeech test set for evaluating performance in scenarios with varying speakers.

\subsection{Implementation Details}

For nearest neighbor retrieval, we construct a FAISS index \cite{Johnson:TBD2021}. We leverage inverted file system and product quantization for quick retrieval from large databases. Keys of examples are stored in fp16 format to conserve memory.

In training our Speaker-Smoothed network, the hidden size is set to 32. We directly use the KeSpeech dev set, training the network for about 4,000 steps. Our model is optimized with Adam, with a learning rate of 3e-4 and a batch size of 32.

During inference, we set the beam search size to 5 and retain other parameters by default
for all settings. For kNN retrieval, we set top $k$ as 32, $\lambda$ as 0.4 and $T$ as 1000 for all experiments.

\begin{table}[tbp]
\centering
\small
\tabcolsep 8pt
\begin{tabular}{l|cc|cc}
\toprule
\multirow{2}{*}{Method}
 & \multicolumn{2}{c|}{Single Speaker} & \multicolumn{2}{c}{Multi Speaker} \\
  & CER    & $\Delta$        & CER         & $\Delta$  \\
\midrule
\textit{Baseline}             & 29.17    & NA        & 36.71         & NA          \\    
\midrule
\multicolumn{5}{l}{In-Domain Setting} \\
\midrule
\textit{Fine-tune}    &\underline{17.29}	&\underline{11.88}	&52.31	&-15.60     \\
\textit{kNN}       &17.92	&11.25	&47.5	&-10.79    \\
\textit{Ours}       &17.74	&11.43	&\underline{19.18}	&\underline{17.53}    \\
\midrule
\multicolumn{5}{l}{All-Domain Setting} \\
\midrule
\textit{Fine-tune}    &18.92	&10.25	&13.05	&23.66     \\
\textit{SAT}         &18.2	&10.97	&14.26	&22.45        \\    
\textit{PAT}         &17.66	&11.51	&12.81	&23.90    \\
\textit{kNN}       &18.11	&11.06	&16.47	&20.24  \\
\textit{Ours}        &17.54	&11.63	&12.3	&24.41   \\
\textit{Fine-tune+Ours}   &\textbf{16.82}	&\textbf{12.35}	&\textbf{12.03} &\textbf{24.68}    \\
\bottomrule
\end{tabular}
\caption{Main Results in two settings. $\Delta$ means the difference from baseline.}
\label{tab:emea-data}
\vspace{-0.3cm} %
\end{table}

\subsection{Main Results}

Table \ref{tab:emea-data} compares the proposed method with other adaptation methods in terms of CER on single and multi speakers. The Whisper baseline results showed that the CER performance was poor in both test sets, with mismatch.

In the in-domain experiment, fine-tuning yielded optimal results for single speaker adaptation. Yet, due to catastrophic forgetting, the multi-speaker test set experienced a CER increase. The kNN method's significant CER decline on the single speaker test set denotes its effectiveness, but rise in the multi-speaker due to mismatching and inability to reject noise. Our Speaker-Smoothed kNN method, albeit slightly weaker than fine-tuning, excelled on the multi-speaker test set.

In the All-domain experiment, fine-tuning underperformed compared to its in-domain variant on the single speaker test set. Still, the multi-speaker test set's CER significantly reduced compared to the baseline, owing to a complete match. We implemented speaker-aware training (SAT) \cite{feature7}, applicable to various ASR models, and PAT \cite{gu23_interspeech}, learning speaker adaptation from the training dataset, for comparison, maintaining the same supervised settings. Both SAT and kNN fell short of fine-tuning on the multi-speaker test set. Our proposed method, however, outperformed on both single and multi-speaker test sets, enhancing further when used continuously on a fine-tuning base.

\begin{figure}[t]
  \centering
  \includegraphics[width=\linewidth]{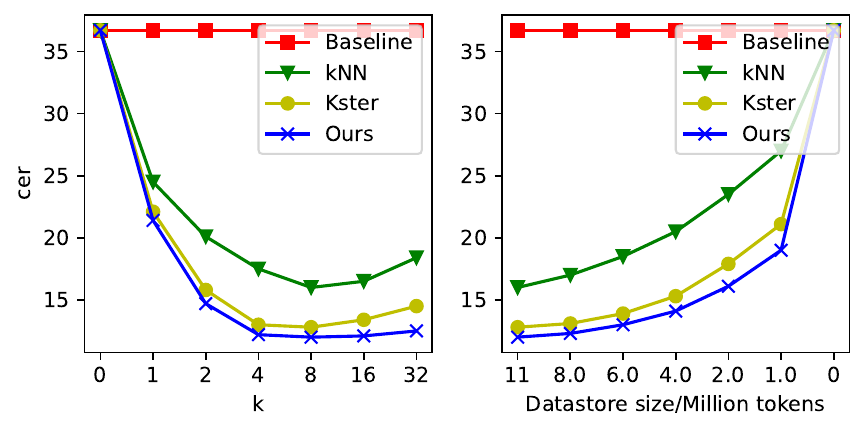}
  \caption{Left side is the CER trend of different methods when use different top k,  right side is the CER trend in different sample datastore size. Both experiment done on the multi-speaker test of all-domain setting. }
  \label{fig:topK}
\vspace{-0.3cm} %
\end{figure}

\subsection{Analysis}

In the all-domain experimental setting with a multi-speaker test set, as shown on the left of Figure \ref{fig:topK}, the top k types of noise increase gradually as k expands, particularly for sparse data. Compared to the direct application of kNN, and Kster \cite{jiang2021learning}, which employ a similar smoothed network without an x-vector, performance diminishes to various extents. As displayed on the right of Figure \ref{fig:topK}, as the datastore contracts, the noise volume incrementally escalates, and our method exhibits a more gradual decline. These two experiments underscore that our Speaker-Smoothed method handles error recalls more effectively.

\subsection{Inference Cost}

The inference expense of kNN memorization retrieval, discussed in previous studies \cite{Du2022NonParametricDA,Dai2023SimpleAS}, may result in reduced speed as the datastore expands. This study explores this on the all-domain setting based on Whisper-medium. With a batch size of 16, we find that the average inference speed is 87.1\% of that of the kNN-free method (given that the weight of the SRE model is relatively light, its consumption was disregarded). In practice, this speed reduction is acceptable as it represents a balancing act between performance and processing time. To achieve superior inference speed, we could potentially replace it with other memorization-retrieval variants \cite{Meng2021FastNN,Dai2023SimpleAS}. This is a prospect for future work.

\section{Conclusion}

Our research presents a novel approach to enhancing E2E speaker adaptation performance through Speaker-Smoothed kNN, notably in situations of limited adaptation data. By utilizing x-vector information, we achieve dynamic adjustment of interpolation ratios, leveraging the similarity in voice characters of training data. We attain considerable improvement, outperforming established techniques like fine-tuning and speaker-aware training in the KeSpeech and MagicData test sets. By demonstrating the compelling application of kNN in ASR speaker adaptation, we pave the way for future inquiries into this promising area.

\bibliographystyle{IEEEtran}
\bibliography{template}

\end{document}